\documentclass[a4paper, 10pt]{article}

\usepackage{graphics}
\usepackage{epsfig,amssymb,euscript,xspace, color}
\usepackage{amsmath,jheppub,mathtools}
\def\bea{\begin{eqnarray}}
\def\eea{\end{eqnarray}}
\def\be{\begin{equation}}
\def\ee{\end{equation}}

\newcommand{\bra}{\langle}
\newcommand{\ket}{\rangle}
\newcommand{\rpl}{r_{\!_+}}
\newcommand{\rmi}{r_{\!_-}}
\newcommand{\Ll}{\mathcal{L}}
\newcommand{\Llo}{\mathcal{L}_{\phi_1}}
\newcommand{\Llt}{\mathcal{L}_{\phi_2}}



\def\cL{{\cal L}}

\def\cS{{\cal S}}

\def\cL{{\cal L}}

\begin{document}

\title{Fast Scrambling of mutual information in Kerr-AdS$_{\textbf{5}}$ }
\author{Vinay Malvimat$^a$, Rohan R. Poojary$^b$}
\affiliation{$^a$Theory Division, Saha Institute of Nuclear Physics, \\
Homi Bhaba National Institute (HBNI), \\
1/AF, Bidhannagar, Kolkata 700064, India. 
\vspace{0.3cm}
\\
$^b$Institute for Theoretical Physics, TU Wien,\\
Wiedner Hauptstrasse 8-10, 1040 Vienna, Austria.  }
\emailAdd{vinaymmp@gmail.com, rpglaznos@gmail.com}

\abstract{ We compute the disruption of mutual information in a TFD state dual to a Kerr black hole with equal angular momenta in $AdS_5$ due to an equatorial shockwave. The shockwave respects the axi-symmetry of the Kerr geometry with specific angular momenta $\mathcal{L}_{\phi_1}$ \& $\mathcal{L}_{\phi_2}$. The sub-systems considered are hemispheres in the $left$ and the $right$ dual CFTs with the equator of the $S^3$ as their boundary. We compute the change in the mutual information by determining the growth of the HRT surface at late times. We find that at late times leading up to the scrambling time the minimum value of the instantaneous Lyapunov index $\lambda_L^{(min)}$ is bounded by $\kappa=\frac{2\pi T_H}{(1-\mu \,\Ll_+)}$ and is  found to be greater than $2\pi T_H$ in certain regimes with $T_H$ and $\mu$ denoting  the black hole's temperature and the horizon angular velocity respectively while $\mathcal{L}_+=\mathcal{L}_{\phi_1}+\mathcal{L}_{\phi_2}$. We also find that for non-extremal geometries the null perturbation obeys $\mathcal{L}_+<\mu^{-1}$ for it to reach the outer horizon from the $AdS$ boundary. The scrambling time at very late times is given by $\kappa\tau_*\approx\log \mathcal{S}$ where $\mathcal{S}$ is the Kerr entropy. We also find that the onset of scrambling is delayed due to a term  proportional to $\log(1-\mu\,\Ll_+)^{-1}$ which is not extensive and does not scale with the  entropy of Kerr black hole.    }

\maketitle


\section{Introduction}
It has been well known in recent years that quantum chaos exhibited by large-$N$ thermal field theories is holographically dual to the near horizon dynamics of black holes \cite{Shenker:2013pqa}. A hallmark of such systems is that they exhibit what is known as fast scrambling $i.e.$ the time taken $t_*$ for any initial $\mathcal{O}(1)$ perturbation to grow is proportional to the degrees of freedom $t_*\sim \log\mathcal{S}$ \cite{Hayden:2007cs,Sekino:2008he}. Black holes in quantum gravity were then conjectured to be among the fastest scramblers. This expectation was validated by Shenker and Stanford for the case of static black holes in AdS by analysing the disruption of the finely tuned  entanglement between the 2 CFTs constituting the TFD state dual to the black hole geometry \cite{Shenker:2013pqa}. This was also followed by the analysis of the \emph{out-of-time-ordered-correlator} (OTOC) in the field theory computed holographically for static BTZ which exhibited an exponential decay
\be
\frac{\bra W(0)V(t)W(0)V(t) \ket}{\bra WW\ket\bra VV \ket}=1-\epsilon e^{\lambda_L t},\hspace{0.3cm}\lambda_L=2\pi T_H
\ee 
where $\epsilon\sim c^{-1}$ and $T_H=\beta^{-1}$ is the temperature of the geometry \cite{Shenker:2014cwa}. The exponential index $\lambda_L$ is what is termed as the Lyapunov index and generally determines how quickly the  scrambling of information takes place. 
Using the analyticity property of the four point  correlators of a thermal QFT in  the complex time plane  it was famously shown by Maldacena, Shenker \& Stanford \cite{Maldacena:2015waa} that the Lyapunov index for time scales leading up to the scrambling time is bounded to be
\be
\lambda_L\leq\frac{2\pi}{\beta}
\label{MSS_bound}
\ee
thus bounding the scrambling time of such systems. The analysis of the 1d solvable strongly coupled SYK-like models  by Maldacena \& Stanford \cite{Maldacena:2016hyu} had revealed similar chaotic behaviour with $\lambda_L=2\pi/\beta$. Here the 1d SYK model possesses a 1d reparametrization symmetry at zero temperature which is spontaneously broken at small temperatures, giving rise to an effective theory of 1d reparametrizations in the IR governed by the Schwarzian action. The holographic dual of such a 1d theory was found by analysing a 2d dilaton gravity theory in $AdS_2$ first analysed by Jackiw and Teitelboim called the JT theory \cite{Jensen:2016pah,Maldacena:2016upp}. This theory reproduced the Schwarzian action as its boundary effective action for scalar perturbations corroborating the behaviour \eqref{MSS_bound} for scalar correlators at its 1d boundary. It was systematically shown that the JT theory effectively describes dynamics of small departures from extremality due to excess mass\footnote{In all these analysis the rest of the black hole charges are held constant.} for a wide class of black holes \cite{Nayak:2018qej,Moitra:2018jqs,Castro:2018ffi,Moitra:2019bub,Ghosh:2019rcj,Castro:2021fhc,Castro:2021csm}. Here, the JT theory is obtained as an effective near horizon gravity theory after dimensionally reducing the higher dimensional Einstein-Hilbert action (with or without cosmological constant) about a near extremal black hole with excess mass. It is also worth noting that chaotic behaviour is observed in holographic systems associated with the dynamics of a string stretched in a black hole back ground\cite{deBoer:2017xdk}, and associated modes have also been identified in this case and  for similar theories involving branes with a Schwarzian effective action being associated with such modes too \cite{Banerjee:2018kwy,Banerjee:2018twd}. This further suggests that the chaotic behaviour associated with fast scrambling is a hallmark of holographic theories of gravity. 
\\\\
The case for black holes with rotation is much more subtle. The analysis of the four point OTOC in BTZ revealed the existence of 2 Lyapunov exponents, $\lambda_L=\frac{2\pi}{\beta_\pm}$, each corresponding to the left and right temperatures of the dual CFT$_2$ of which one survives the extremal limit \cite{Poojary:2018esz,Stikonas:2018ane,Jahnke:2019gxr}. The generalization of the arguments leading up to the  MSS bound for the case of thermal QFT with chemical potential $\mu$ revealed that $\lambda_L$ obeys a bound of the form
\be
\lambda_L\leq\frac{2\pi T_H}{(1-\mu/\mu_c)}
\label{Halder}
\ee
with $\mu_c$ being the maximum possible value for the chemical potential \cite{Halder:2019ric}. This bound is saturated by the larger of the 2 temperatures of the CFT$_2$. Similar behaviour was also observed for the case of string stretched in rotating BTZ and $AdS_4$ Kerr geometries, in the former case an effective theory was also obtained \cite{Banerjee:2019vff}.
\\\\
``Pole skipping'' is yet another phenomena that is closely associated to chaotic phenomena in fast scrambling systems where retarded energy 2pt functions in frequency space have been observed to skip poles at frequency $\omega_L=i\lambda_L$ and wave-number $k=\pm i v_B$, $v_B$ being the butterfly velocity \cite{Grozdanov:2017ajz,Blake:2017ris}. Static branes in a black hole background in $AdS$ have been shown to exhibit pole skipping at $\lambda_L=2\pi T_H$ with an effective theory of hydrodynamics associated with them \cite{Blake:2017ris,Blake:2018leo}. For the case of rotating black holes in BTZ pole skipping was observed for the 2 temperatures of the dual CFT$_2$ \cite{Liu:2020yaf}. It is also worth noting that the pole skipping analysis by investigating ingoing modes of the graviton in rotating Kerr $AdS_4$ revealed $\lambda_L=2\pi T_H$ \cite{Blake:2021wqj}.
\\\\
However the scrambling time for the 4pt OTOC in a CFT dual to a rotating BTZ revealed that the long time averaged exponent is $|\lambda_L|=\frac{2\pi}{\beta}$ \cite{Mezei:2019dfv,Craps:2020ahu}, while the growth in time of the OTOC exhibited a sawtooth like pattern with $\lambda_L^{(max)}$ being the larger of the 2 temperatures of the CFT$_2$. These were also reproduced from the CFT$_2$ perspective \cite{Craps:2021bmz} assuming vacuum block domination.
\\\\
The mutual information $I(A:B)$ between the 2 boundary CFTs represented in the TFD description can serve as a better diagnostic of chaotic phenomena \cite{Wolf:2007tdq}. Here $A$ and $B$ are large enough subsystems in the left and right boundary CFTs respectively. The disruption of $I(A:B)$ due to an $O(G_N)$ perturbation at late times can be computed using the RT or the HRT prescription \cite{Ryu:2006bv,Hubeny:2007xt,Shenker:2013pqa}. The analysis in \cite{Shenker:2013pqa} realises the late time effect of such a perturbation  by a shockwave in the black hole geometry with the rate of disruption of $I(A:B)$, thus determining $\lambda_L$. A crucial physical insight of this analysis was  that the $\lambda_L$  turns out to be given by the exponent of the blueshift suffered at the outer horizon by an in-falling null particle  released from the boundary. For the case of rotating shockwaves in rotating BTZ it was shown that this blueshift is given by
\be
E=E_0e^{\kappa t_0},\hspace{0.3cm} \kappa=\frac{2\pi T_H}{(1-\mu\,\Ll)}
\label{rot_blueshift}
\ee
where $\mu=\rmi/\rpl$ is the chemical potential and $\Ll$ is the angular momentum of the shockwave per unit of its energy \cite{Malvimat:2021itk}. It was also shown by an analysis similar to \cite{Shenker:2013pqa} that in certain cases $\lambda_L>2\pi T_H$ and such a $\lambda_L$ also determines the scrambling time\footnote{Here we explicitly see that $I(A:B)$ is a better diagnostic of chaotic phenomena \cite{Wolf:2007tdq} than 4pt OTOC where the later sees an $|\lambda_L|=2\pi/\beta$ for large times \cite{Mezei:2019dfv}. }.
Recently the disruption of mutual information due to rotating shockwaves was analysed for Kerr black holes in $AdS_4$ \cite{Malvimat:2022oue}. Here the $A$ \& $B$ subsystems were chosen to be the hemispheres of the left and right CFTs respectively such that they respect the axi-symmetry of the geometry. Here too we find that  the blueshift suffered by a rotating shockwave released at time $t_0$ from the boundary is given by \eqref{rot_blueshift}. In this case we clearly find that the instantaneous $\lambda_L^{(min)}$ leading up to the scrambling time can be greater than $2\pi T_H$ while the late time average $\lambda_L$ that determines the scrambling time is given by 
\be
\lambda_L=\kappa=\frac{2\pi T_H}{(1-\mu\,\Ll)}
\label{rot_lambda}
\ee 
This is aslo not in contradiction with the result  obtained in \cite{Blake:2021wqj} mentioned before as it investigates pole skipping by looking at non-rotating in-going graviton modes at the horizon $i.e.$ $\mathcal{L}=0$; the work in \cite{Malvimat:2022oue} predicts that similar analysis with rotating modes would corroborate the above results.
In this paper we provide further evidence of such phenomena in Kerr black holes in $AdS_5$ with equal angular momentum. We analyse the disruption of $I(A:B)$ where $A$ \& $B$ are hemispheres in the left and right boundary CFT$_4$s respectively $s.t.$ they respect the axi-symmetry of the geometry. The disruption is caused due to a rotating shockwave with angular momenta per unit energy $\Llo$ and $\Llt$ about the 2 axi-symmetric angular directions of the geometry.
The equal angular momentum Kerr $AdS_5$ possesses an $S^{2}$ worth of isometry which makes the analysis easier. The  paper is organized as follows: we first briefly review the Kerr geometry with equal angular momentum in section 2. 
In section 3 we construct the Dray-'t Hooft solution for rotating shockwaves. 
This we do by constructing the necessary Kruskal coordinates suitable for expressing the Dray-'t Hooft solution. 
In the process we also obtain the blue shift suffered by  such a shockwave released from the boundary in reaching the outer horizon. We also briefly comment about the turning point analysis required for rotating null geodesics to reach the horizon from the boundary and show that the exponent of the blueshift does not diverge. 
In section 4 we determine the change in the area of the extremal surface following the computation in \cite{Leichenauer:2014nxa}, due to such a shockwave and plot its instantaneous $\lambda_L^{(min)}$ for various values of shockwave angular momenta along with the respective blueshift exponents $\kappa$ against $\rmi/\rpl$ in Fig-\ref{F_cH_c_kappa}. We also compute the scrambling time $i.e.$ time taken for $I(A:B)$ to go to zero and find that the late time $\lambda_L$ is given by \eqref{rot_lambda} where $\Ll=\Ll_+=\Llo+\Llt$. 
\section{Kerr in AdS$_\textbf{5}$ with $J_{\phi_1}=J_{\phi_2}$}
In this section we briefly introduce the geometry of an  $AdS_5$ Kerr black hole  with equal angular momentum. The asymptotic boundary of $AdS_5$ is a $\mathbb{R}\times S^3$ however a Kerr geometry in the bulk does not have the rich isometries of $S^3$ ($SU(2)\times SU(2)$). Choosing the 2 angular momenta of the Kerr geometry to be equal $J_{\phi_1}=J_{\phi_2}=J$ preserves an $S^2$ fibration of $S^3$, therefore the geometry possesses an $SO(3)\subset SU(2)\times SU(2)$ isometry. This allows for a much simpler expression for the metric in terms of the line element squared. 
The equal angular momentum Kerr black hole in $AdS_5$ in Boyer-Lindquist coordinates is
\bea
ds^2&=&\frac{r^2dr^2}{(r^2+a^2)\Delta(r)}\,-\,\frac{1}{\Xi}\Delta(r)e^{U_2-U_1}dt^2\,+\,e^{-U_1}d\Omega_2^2\,+\,e^{-U_2}\left(\sigma^3+A\right)^2,\cr&&\cr&&\cr
\hspace{-2cm}{\rm where}\hspace{3cm}&&\Delta(r)=1+\frac{r^2}{l^2}-\frac{2mr^2}{(r^2+a^2)^2},\hspace{0.3cm}
\Xi=1-\frac{a^2}{l^2},\cr&&\cr
&&e^{-U_2}=\frac{r^2+a^2}{4\,\Xi}+\frac{2ma^2}{2\,\Xi^2(r^2+a^2)},\hspace{0.3cm}e^{-U_1}=\frac{r^2+a^2}{4\,\Xi},\cr&&\cr
&&A=\frac{a}{2\,\Xi}\left(\frac{r^2+a^2}{l^2}-\frac{m}{r^2+a^2}\right)e^{U_2}dt	
\label{Kerr_Boyer_lindquist}
\eea
The angular line elements are
\be
\sigma^3=d\psi+\cos \theta\,d\phi,\hspace{0.4cm}d\Omega^2_2=d\theta^2+\sin^2\theta\,d\phi^2
\ee
where the $S^2$ symmetry is manifest. The polar coordinates range from $\theta\in \{0,\pi\}$ ,  $\phi\in\{0,2\pi\}$ and $\psi\in\{0,4\pi\}$\footnote{The toric coordinates describing the $S^3$ are $\eta=\theta/2$ and $\phi_1=(\psi-\phi)/2$, $\phi_2=(\psi+\phi)/2$ where \\$d\Omega_3^2=d\eta^2+\sin^2\theta \,d\phi_1+\cos^2\theta \,d\phi_2$.   }.
The axi-symmetry directions of the black hole are $\phi_1=(\psi-\phi)/2$ and $\phi_2=(\psi+\phi)/2$. The black hole is defined by the mass parameter $m$ and the angular momentum parameter $a$. 
The ADM mass and 2 equal angular momenta are \cite{Skenderis:2006uy}
\be
M=M_C+\frac{2\pi^2 m(3+\frac{a^2}{l^2})}{\Xi^2},\hspace{0.3cm}J=\frac{8\pi^2 m a}{\Xi^2},\hspace{0.3cm}
M_C=\frac{3\pi^2l^2}{4}
\ee
Here $M_C$ is the Casimir energy associated with $S^3$ and is independent of the black hole parameters.
The entropy and temperature is
\be
\mathcal{S}=\frac{\pi^2(\rpl^2+a^2)^2}{2G_N\rpl \Xi^2},\hspace{0.5cm}T_H=	\frac{\rpl^2-a^2+\frac{2a^2}{l^2}}{2\pi\rpl(\rpl^2+a^2)}
\ee
where $l$ is the radius of $AdS_5$ an $G_N$ is the Newton's constant.
The Boyer-Lindquist coordinates used above are adapted from that of flat space, in asymptotically $AdS_5$ space these have a non-zero angular velocity at the boundary about $\phi_1$ and $\phi_2$. It turns out that the correct horizon angular velocity about both these directions after cancelling the relative angular velocity  of the boundary is
\be
\Omega=\Omega_{\phi_1}=\Omega_{\phi_2}=\frac{a\left(1+\frac{\rpl^2}{l^2}\right)}{\rpl^2+a^2}=\mu
\label{mu}
\ee
which
plays the role of chemical potential $\mu$ for both the angular momenta and is relevant for the thermodynamics of the black hole. The reader is referred to \cite{Skenderis:2006uy} for a detailed derivation of asymptotic charges relevant for thermodynamics for generic Kerr black holes in $AdS_5$.
\section{Dray-'t Hooft solution}
In this section we construct the Dray-'t Hooft solution corresponding to a shockwave along the equator $\theta=\pi/2$ such that the shockwave and thus the resulting geometry obeys the axi-symmetries of the Kerr black hole. By symmetry of the Kerr solution it can be shown that there always exists a geodesic (massive and massless) which stays restricted to the equatorial plane. As the Dray-'t Hooft solution requires well defined Kruskal coordinates smooth at the outer horizon we first construct such coordinates by working out the null geodesics with angular momenta per unit energy $\Ll_{\phi_1}$ \& $\Ll_{\phi_2}$ about the two symmetric angular coordinates of the Kerr black hole. We then show systematically that the blueshift suffered by such a shockwave is obtained by demanding that the associated in-out going Kruskal coordinates be affine at the outer horizon. The Dray-'t Hooft solution is then obtained by utilizing these Kruskal coordinates and  determining the shift in one of them as a function of time $t_0\gg\beta$ when the perturbation was realised from the left AdS boundary. 
\subsection{Rotating Kruskal Coordinates}
We begin with null geodesics parametrized by the constants
\be
\xi^2=0,\hspace{0.3cm}g_{\mu\nu}\xi^\mu \zeta^\nu_t=\mathcal{E}=1,\hspace{0.3cm}
g_{\mu\nu}\xi^\mu\zeta^\nu_{\phi_1}=\mathcal{L}_{\phi_1},\hspace{0.3cm}g_{\mu\nu}\xi^\mu\zeta^\nu_{\phi_2}=\mathcal{L}_{\phi_2}
\label{geodesic_eq}
\ee
where $\zeta_{\phi_1},\zeta_{\phi_2}$ and $\zeta_t$ are the symmetry generators with respect to the stationary boundary observer. In terms of the Boyer-Lindquist coordinates these are 
\bea
&&\zeta_{\phi_1}=\partial_{\phi_1}, \hspace{0.3cm}\zeta_{\phi_2}=\partial_{\phi_2}, \hspace{0.3cm}\zeta_t=\partial_t-\frac{2a}{l^2}\partial_\psi\cr&&\cr
&&{\rm where}\hspace{0.3cm}\phi_1=\frac{\psi-\phi}{2},\hspace{0.3cm}\phi_2=\frac{\psi+\phi}{2}
\eea
There is a further constant that needs to be specified which corresponds to the angular momentum in the $\theta$ direction. This is done by specifying  the Carter's constant $\mathcal{Q}$ which is defined by using a higher form of symmetry present in rotating geometries called the Killing-Yano tensor $f_{\mu\nu}$. The Killing-Yano tensor can be used to write down a higher form generalization of the Killing vector called the Killing tensor which satisfies
\be
\nabla_\rho K_{\mu\nu}=0,\,\,\,K_{\mu\nu}=K_{\nu\mu},\,\,\,K_{\mu\nu}=-f_\mu^{\,\,\,\sigma}f_{\sigma\nu}
\ee
The Carter's constant is then defined as
\be
\xi^\mu\xi^\nu K_{\mu\nu}=\mathcal{Q}
\ee  
The Killing tensor for generic Kerr $AdS_5$ is worked out in \cite{Kunduri:2005fq}. We make use of it in our coordinates for the case of equal angular momentum. Imposing that the geodesic stays in the equatorial plane $i.e.$ $\dot{\theta}=\xi^\theta=0$ implies holding $\mathcal{Q}$ constant at
\be
\mathcal{Q}=\Xi\,(2(\Llo^2+\Llt^2)-l^2).
\label{Equatorial_Q}
\ee
The precise form of the vector field $\xi^\mu\partial_\mu$ along the null geodesic parametrized by the above constants is cumbersome and can be worked out easily. The null pairs  corresponding to in going and out going geodesics can be obtained by reversing the signs of the angular coordinates, we denote these as $\xi^\pm$ and use their dual one-forms to write the metric at $\theta=\pi/2$.
\be
ds^2_{\theta=\frac{\pi}{2}}=F(\xi^+_\mu dx^\mu)(\xi^-_\nu dx^\nu)+h_\phi(d\phi+h_\tau d\tau)^2 +h_z(dz+g_\phi d\phi+A_\tau d\tau)^2
\ee
where
\be
\tau=\left(1-\frac{a}{l^2}(\Ll_{\phi_1}+\Ll_{\phi_2})\right) t + \frac{\Ll_{\phi_1}-\Ll_{\phi_2}}{2}\phi-\frac{\Ll_{\phi_1}+\Ll_{\phi_2}}{2}\psi
\label{tau_to_t_phi_psi}
\ee
Here the $\tau$ naturally occurs as the component of the one-form in $\xi^\pm_\mu dx^\mu$ in the non-radial direction. The rest of the functions in the line elements written above can be obtained uniquely by completing the squares. $z$ is a linear function of $\tau$ and $\psi$ which we elaborate below. Although the forms of these functions are important, their precise expressions are cumbersome and hence it serves little purpose to explicitly give them here. Their behaviour close to the horizon would be significant for our analysis as we shall see further. The $\tau$ coordinate in terms of the Kruskal coordinates reads
\be
d\tau\sim\frac{UdV-VdU}{UV}
\ee
therefore the coefficient of the $d\tau$ term in the line elements better vanish at the horizons.
The function $h_\tau\sim\mathcal{O}(r-\rpl)$, therefore the $\phi$ coordinate need not be redefined.
However the $\psi$ coordinate is shifted to $z$ to demand that $A_\tau\sim\mathcal{O}(r-\rpl)$.
\bea
\psi&=&\eta\, z-\frac{2a(a^2-l^2)}{l^2(a^2+\rpl^2)-a(\rpl^2+l^2)(\Ll_{\phi_1}+\Ll_{\phi_2})}\tau
\label{psi_to_z_tau}
\eea
The coefficient of $z$ denoted by $\eta$ in the above expression is fixed by demanding that the horizon area remain the same in the coordinates $\{\theta,\phi,z\}$ and that $z$ and $\psi$ have the same domain $\{0,2\pi\}$.
\be
\eta=\frac{1-\tfrac{a}{l^2}(\Ll_{\phi_1}+\Ll_{\phi_2})}{1-\mu (\Ll_{\phi_1}+\Ll_{\phi_2})}, \hspace{0.8cm}{\rm where\,\,\,\,} \mu=\Omega
\label{psi_to_z_coeff}
\ee
The $z$ coordinate is basically a co-rotating coordinate close to the horizon. Light cone coordinates can be constructed by defining 
\bea
&&du=\xi^+_\mu dx^\mu \implies u=r_*-\tau\cr&&\cr
&&dv=\xi^-_\mu dx^\mu \implies v=r_*+\tau
\label{LC_coordinates}
\eea
where the radial tortoise coordinate is defined by integrating $dr_*=\xi_rdr$ and choosing the boundary at $r_*=0$. We can see that these $\{u,v\}$ coordinates are not smooth at the horizons; for instance for the vector field $\chi$ that generates translations along $v$ $i.e.$ 
$\chi= \partial_u$ we find $\chi\cdot\nabla\chi=\mathcal{K}\chi$ where
 \be
\mathcal{K}=\frac{1}{2}\xi^+\cdot\partial F 
 \ee
 Here the non-vanishing of $\mathcal{K}$ at the outer horizon signals the non-smoothness of the $\{u,v\}$ coordinates. 
One can therefore define the smooth Kruskal coordinates as
\be
U=-e^{\kappa(r_*-\tau)},\,\,V=e^{\kappa(r_*+\tau)},\hspace{0.7cm}{\rm where} \,\,\kappa=\mathcal{K}\vert_{\rpl}
\label{Kruskal_coordinates_equator}
\ee
Rewriting the line element along the equator in terms of $\{U,V\}$ we find
\be
ds^2_{\theta=\frac{\pi}{2}}=\frac{F}{\kappa^2 UV}dUdV + h_z(dz+g_\phi d\phi+A_\tau d\tau)^2 +h_\phi(d\phi+h_\tau d\tau)^2 
\label{Kruskal_equator_netric}
\ee
where the value of $\kappa$ is
\bea
&&\kappa=\frac{\kappa_0}{(1-\mu(\Ll_{\phi_1}+\Ll_{\phi_2}))},\,\,\,\,{\rm where}\,\,\mu=\Omega\,\,\,\,\,\&\,\,\,\,\,\kappa\vert_{(\Llo+\Llt)\rightarrow0}=\kappa_0=\frac{2\pi}{\beta}=2\pi T_H
\label{kappa_0}
\eea
Here we denote $\kappa_0$ to be $2\pi$ times the black hole temperature.
\subsection{Full Coordinates}
We would like to extend these coordinates along the angular coordinate $\theta$ so as to describe the full black hole geometry. Therefore we first write down the null geodesics $\xi^\mu\partial_\mu$ for unit energy with charges $\Llo,\Llt$ and arbitrary $\mathcal{Q}$ with its $\theta$ component. The in-out going vector fields have opposite signs for their $\theta$ component. Next we demand that their dual one-forms be exact 
\be
\xi^+_\mu dx^\mu =du,\hspace{0.3cm}\xi^+_\mu dx^\mu =dv
\ee
This implies 
\be
\partial_\theta\xi^\pm_r=\partial_r\xi^\pm_\theta=0,\hspace{0.4cm}\implies\mathcal{Q}=const 
\ee
We then choose the value of $\mathcal{Q}$ to be \eqref{Equatorial_Q} such that the family of geodesics that start at the equator stay at the equator. Note that this implies $\dot{\theta}=\xi^\theta=0$ only at $\theta=\pi/2$. Thus the whole metric can be written as
\be
ds^2=\frac{F}{\kappa^2 UV}dUdV + h_z(dz+g_\phi d\phi+A_\tau d\tau)^2 +h_\phi(d\phi+h_\tau d\tau)^2  + h_\theta(d\theta+g_\tau d\tau)^2
\label{metric_kruskal_L}
\ee
where we use the same functions as in the previous sub-section to denote the non-trivial $\theta$ dependence. The line elements in the above metric are obtained uniquely by completing squares as before. We also note that $g_\tau\sim\mathcal{O}(r-\rpl)$ implying that $\theta$ need not be redefined as expected. Note that the redefinition of $\psi$ to $z$ is independent of $\theta$.
The Kruskal coordinates can now be defined at any value of $\theta$  to be
\be
U=-e^{\kappa(r_*-\tau-\tilde{g}(\theta))},\,\,V=e^{\kappa(r_*+\tau+\tilde{g}(\theta))},\hspace{0.7cm}{\rm where} \,\,\kappa=\mathcal{K}\vert_{\rpl}
\label{Kruskal_coordinates_equator}
\ee
with an extra $\theta$ dependence $s.t.$ $\tilde{g}(\pi/2)=0$.  $\mathcal{K}$ is similarly defined as
\be
\mathcal{K}=\tfrac{1}{2}\xi^+\cdot\partial F
\ee
The extra $\theta$ dependence is absorbed into $F$ however $\kappa$ is independent of $\theta$ and is still given by
\be
\kappa=\frac{\kappa_0}{(1-\mu(\Ll_{\phi_1}+\Ll_{\phi_2}))},\hspace{0.7cm}{\rm where}\,\,\,\, \kappa_0=\frac{2\pi}{\beta},\,\,\mu=\Omega
\label{kappa}
\ee
The above value of $\kappa$ as in \cite{Malvimat:2021itk,Malvimat:2022oue} suggests an upper bound for the angular momentum of the shock-wave to be $\mathcal{L}<\mu^{-1}=\Omega_+^{-1}$. Here $\mu$ is the same along $\phi_1$ and $\phi_2$. 
\\\\
The functions $A_\tau$, $h_\tau$ and $g_\tau$ vanish at the future or the past outer horizon. The redefinition of $\psi\rightarrow z$ is precisely dictated by this condition. Note that $\theta$ and $\phi$ do not need to be redefined. This requirement would be crucial for writing down the Dray-'t Hooft solutions in response to a shockwave. 
\\\\
The exact functional forms of the functions $\{F,h_z,g_\phi,A_\tau,h_\phi,h_\tau,h_\theta,g_\tau\}$ in the metric \eqref{metric_kruskal_L} are complicated and can be arrived at by following the above procedure after obtaining the in-out going null geodesics \eqref{geodesic_eq}. We note the expansion of the function $F$ at the outer horizon
\be
F(r,\theta)=(r-\rpl)F'(\rpl)+\tfrac{1}{2}(r-\rpl)^2F''(\rpl,\theta)+\dots
\ee 
where we write
\be
(r-\rpl)F'(\rpl)=-\mathbb{A}UV
\label{r_to_UV}
\ee
as the exteriors have either $U$ or $V$ negative. Note that $F'$ is independent of $\theta$. Here $\mathbb{A}$ is some proportionality constant which we assume does not scale with the size or the entropy of the black hole. It would be useful to write $F/UV$ as an expansion in $UV$ using the above expansion
\be
\frac{F}{\mathbb{A}UV}=-1-\frac{\mathbb{A}UV}{2}\frac{F''}{F'^2}.
\label{F_to_UV}
\ee
We will find that such an expansion would be useful for analysing the smoothness condition across the shockwave.
\subsection{Turning point} 
\begin{figure}
	\centering
	\includegraphics[scale=0.17]{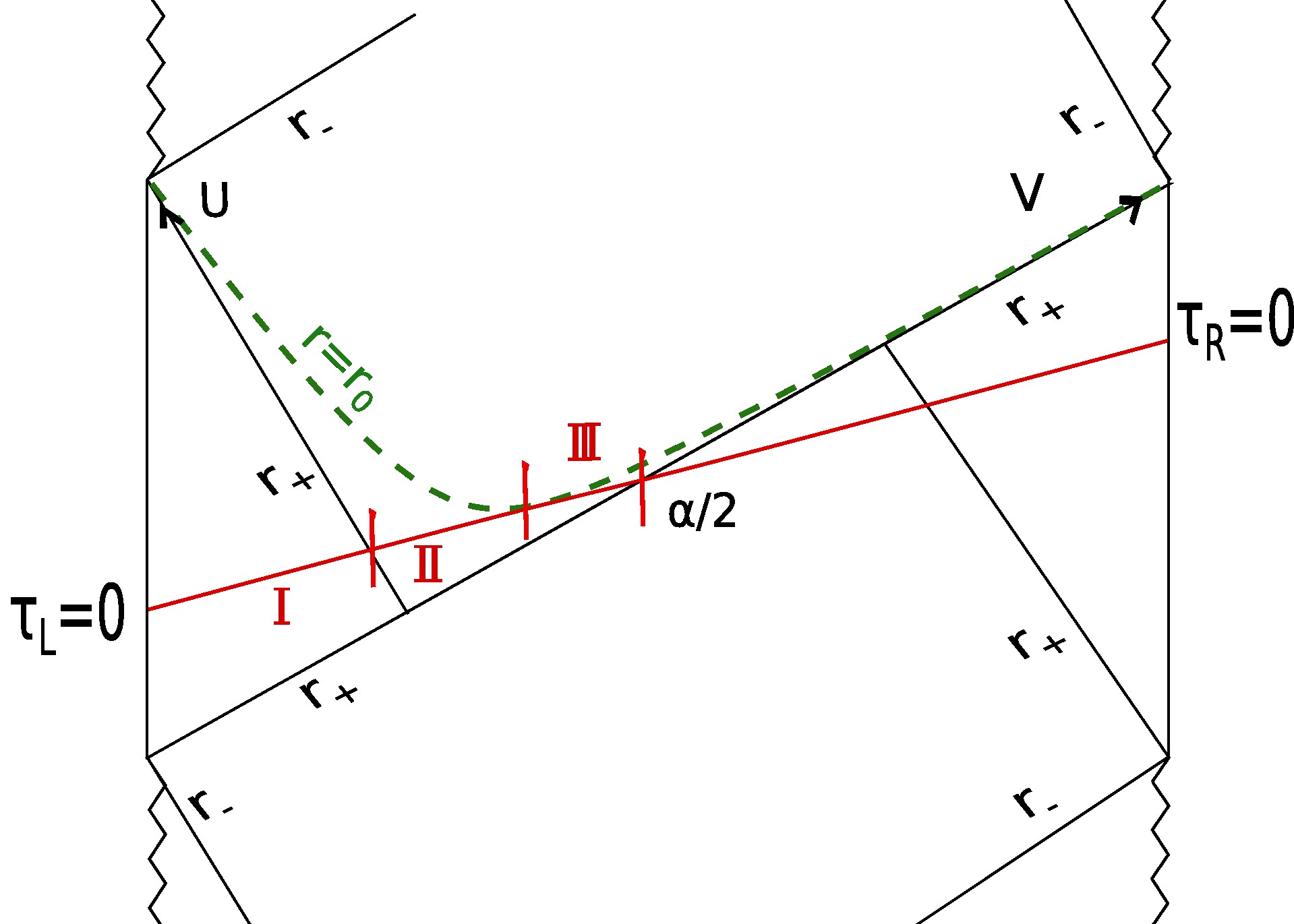}
	\caption{The extremal surface (Red) at $\tau=0$ in the Kruskal extension of Kerr spacetime divided into three segments denoted as I, II and III. The shift in $V$ coordinate at $U=0$ is $\alpha$. Note, the lengths II and III are the same (the diagram is skewed to enlarge the sections) and so are the lengths along the red curve on the either sides of $U=0\,\,\&\,\,V=\alpha/2$. }
\label{KerrExt}
\end{figure}

We would be imagining the perturbation to begin at some point on the (left) AdS boundary and follow a null geodesic towards the outer horizon along $\theta=\pi/2$ Fig-\ref{KerrExt}. This is consistent as $\xi^\theta(r,\theta=\pi/2)=0$.  Not all null perturbations would be able to reach the outer horizon as the geometry is rotating and further the geodesics have a angular momenta about both the symmetric directions. In order to determine whether a null geodesic reaches at least the outer horizon we need to look at the zeros of the radial component of the geodesic vector field \eqref{geodesic_eq}  as it represents the change $\dot{r}$ along the null trajectory. An existence of a zero (of $\xi^r$) outside the outer horizon would indicate that the null perturbation would reverse direction and start moving away from the black hole\footnote{Complicated motions like orbiting the black hole would also be possible, however we would only be interested if the zero exists outside of the horizon and not the fate of the geodesic after it reaches the turning point.} and therefore is called the turning point. The turning point analysis in general is quite  complicated  however for our analysis we only require a few qualitative properties describing the behaviour of turning point as a function of the shockwave's angular momenta $\Llo$ and $\Llt$, the black hole's ratio of inner and outer horizon $\mu_T=\rmi/\rpl$ and its outer radius $\rpl$. We choose to work with 
\be
\Ll_\pm=\Llo\pm\Llt
\label{L_pm}
\ee  
as only $\Ll_+$ determines the blueshift suffered by the null geodesic. We define $\Ll_\pm$ in terms of the inverse of the horizon velocity $\mu$ \eqref{mu}
\be
\Ll_\pm=\epsilon_\pm \mu^{-1}
\label{L_pm_mu}
\ee
We plot $\xi^r(r)$ for $\mu_T\in\{0,1\}$, and $\epsilon_\pm\in\{0,1\}$ for different values of $\rpl$ ranging from very small (compared to AdS radius $l$) to $l/\sqrt{2}$ as this is the largest value of the outer horizon for which one can have an extremal Kerr black hole in $AdS_5$. We find the following behaviour for the largest zero $r_{tp}$:
\begin{itemize}
\item[1] For $\epsilon_+=1$ $i.e.$ $\Ll_+=\mu^{-1}$  we find that $r_{tp}>\rpl$ for non extremal black holes and $r_{tp}\rightarrow\rpl$ as $\mu_T\rightarrow 1$. This holds true for any value of $\Ll_-$ and $\rpl$. Thus we find that the blueshift exponent $\kappa$ \eqref{kappa} does not blow up for non-extremal geometries.
\item[2] As $\epsilon_+$ is reduced from $1$ at some point we do see that $r_{tp}\leq\rpl$ thus indicating that the geodesic can cross the horizon. This limiting value of $\epsilon_+$ is difficult to determine analytically and does depend on $\mu_T$, $\rpl$ and $\epsilon_-$. In general we find that if $r_{tp}>\rpl$ for a certain value of $\epsilon_+$, then increasing $\epsilon_-$ from $0$ to $1$ would further increase the value of $r_{tp}$ $i.e.$ making it difficult for the geodesic to reach the outer horizon.
\item[3] For $\epsilon_+=0$ there are no zeros outside the horizon for $\epsilon_-=0$. For a certain window of $\mu_T\in\{0,\mu_T'\}$ (where $\mu_T'<1$) we find an $r_{tp}>\rpl$ after a certain value of $\epsilon_-$ as it is increased from $0$ to $1$. Here $\mu_T'$ is sufficiently far fro extremality and depends in general on all the other parameters. For $\mu>\mu_T'$ changing $\epsilon_-$ between $\{0,1\}$ has no effect on the turning point and there exist no $r_{tp}>\rpl$.    
\label{tp_1}
\end{itemize}
Generally one analyses the turning points by studying the solution to $\xi^r=0$. However in this case the behaviour of the zeros with regards to $\epsilon_\pm,\mu_T$ and $\rpl$ are too cumbersome and we leave a more complete analysis of the properties of the geodesics in rotating $AdS$ black hole geometries for future investigations. 
Furthermore, the first point mentioned above would be the only characteristic of the null geodesics of direct relevance for us as it clearly rules out the possibility of the exponent of the blueshift $\kappa$ \eqref{kappa} blowing up for a non extremal configuration.  
\\\\
Given the results of \cite{Halder:2019ric} we expect Lyapunov index we wish to compute to be bounded by \eqref{Halder}. As we would later see that since the Lyapunov index turns out to be dictated by the exponent of the blueshift suffered along the null in-falling geodesic we expect $r_{tp}>\rpl$ for $\cL_+> \mu_c^{-1}$ where $\mu_c$ is the horizon velocity for an extremal black hole with the same ADM mass. This would require a further numerical analysis of the turning point and we leave the this thorough analysis for the near future.
\subsubsection{Superradiance}
It is worthwhile to comment about Superradiance due to an in-falling rotating perturbation into the Kerr black hole under consideration. Superradiance can be understood as a classical phenomena\footnote{For discussion regarding the quantum effects of superradiance refer to \cite{Brito:2015oca}.} wherein the particle falling into a Kerr space-time with an angular momentum enters the ergo-region surrounding the black hole horizon and exists with an energy greater than which it began at an early time. This process is also called the Penrose process whereby one can extract the rotational energy of a Kerr black hole. For a detailed review describing the current understanding of numerous such cases, we refer the reader to the extensive lecture notes of \cite{Brito:2015oca} which is continuously updated\footnote{We thank Indranil Halder for this excellent reference.}. This process can be seen as a simple consequence of the first law and the area increase theorem (second law) for black hole dynamics. Let us assume that the particle has an angular momentum per unit of its energy $\Ll$. The change in the mass $M$, entropy $S$ and angular momentum $J$ of the black hole due to this interaction would be
\bea
&&\hspace{-2cm}\delta M=T_H\,\delta\mathcal{S}+\mu\,\delta J\cr&&\cr
{\rm for}\hspace{0.4cm}\delta J=\Ll\delta M&\implies&\delta M=\frac{T_H}{(1-\mu\,\Ll)}\delta \mathcal{S}
\label{superradiace_first_law}
\eea
The area increase theorem for black holes implies that $\delta\mathcal{S}>0$ in any such process. Therefore for the particle to exit the ergo-region with greater energy the change in the black hole's mass must be $\delta M<0$, which is possible only if $\mathcal{L}>\mu^{-1}$.  For our case of Kerr black hole in $AdS_5$ with $\mu=\mu_1=\mu_2$ we simply have $\delta J=\delta J_1+\delta J_2$ and $\Ll=\Ll_+$ defined above.
In other words the perturbation ought to have a specific angular momentum $\Ll$ such that its trajectory enters the ergo-region and leaves it $i.e.$ doesn't fall into the black hole. This is consistent with the above turning point analysis where we found that the necessary condition for the geodesic to fall into the non-extremal geometry is $\Ll_+<\mu^{-1}$. For specific angular momentum greater than this the turning point lies outside the horizon, thus the null particle reverses its direction in terms of the radial coordinate and depending on the geodesic may either escape to a region far from the black hole or simple be trapped in an orbit around it. In the former case we get to see the Penrose process while in the later the null trajectory gets trapped. Either of these scenarios would not be of concern to us as we would be interested in only those values of $\Ll_+$ wherein the geodesic makes it past the outer horizon.
\subsection{Shockwaves along the equator}
The Dray-'t Hooft solution is given by
\be
V\rightarrow\widetilde{V}=V+\alpha\Theta(U)f
\label{V_shift_0}
\ee
where $f$ in general is function of the spatial coordinates $\{\theta,z,\phi\}$ transverse to the shockwaves. For the sake of simplicity we would send shockwaves with axi-symmetry of the geometry $i.e.$ the shockwave would be symmetric in the $z$ and $\phi$ directions and be localized  on the equatorial plane $\theta=\pi/2$. Thus $f$ would only depend on $\theta$.
\\\\
The metric in $\{U,V,\theta,\phi,z\}$ coordinates can be obtained by first noting $d\tau$ in terms of Kruskal coordinates
\be
d\tau=\frac{1}{2\kappa \,UV}(UdV-VdU)-\widetilde{g}'d\theta	
\ee
yielding the Kerr metric in Kruskal coordinates as
\bea
ds^2_{}&=&\frac{F}{\kappa^2 UV}dUdV + h_z\left(dz+g_\phi d\phi+\frac{A_\tau}{2\kappa\,UV}(UdV-VdU) +A_\tau\widetilde{g}'d\theta \right)^2 +\cr&&\cr
&&+h_\phi\left(d\phi+\frac{h_\tau}{2\kappa\,UV}(UdV-VdU) +h_\tau\widetilde{g}'d\theta\right)^2  + h_\theta\left(d\theta+\frac{g_\tau}{2\kappa\,UV}(UdV-VdU) +g_\tau\widetilde{g}'d\theta\right)^2\cr&&
\label{metric_kruskal_L_0}
\eea
Note that along and across the horizon the above metric is smooth, this is guaranteed by the behaviour of the functions $h_\tau,g_\tau\,\&\,A_\tau$ all vanishing as $\mathcal{O}(r-\rpl)$ at the outer horizon. The Dray-'t Hooft solution for a shockwave parametrized by $U=0$ can now be written following the general arguments presented in \cite{Sfetsos:1994xa} as
\be
ds^2\rightarrow d\widetilde{s}^2-\frac{F}{\kappa^2UV}\delta(U)f(\theta)dU^2
\label{shockwave_sol_0}
\ee
The above form of the metric implies that the Einstein's equation
\be
R_{\mu\nu}-\frac{1}{2}Rg_{\mu\nu}-\Lambda g_{\mu\nu}=R_{\mu\nu}+\frac{4}{l^2}g_{\mu\nu}=-8\pi G_N T_{\mu\nu}
\ee
with the $rhs$ corresponding to that of a stress-tensor due to shockwave with momentum $p^V$ along the $V$ coordinate. The above equation then reduces to that of source equation for the response function $f(\theta)$ in the transverse direction
\be
\alpha\mathcal{D}_{\theta}\,f=-8\pi G_N T_{UU},\,\, T_{UU}=p^V\left(\frac{F^2}{\kappa^2UV}\right)_{U_0}\delta(\theta-\pi/2)
\label{backreaction_diff_axi_symmetric}
\ee
where $\mathcal{D}_\theta$ is a differential operator obtained by evaluating the Einstein's equation at $U=0$. Note, that its is important here that the metric \eqref{metric_kruskal_L_0}  be smooth at $U=0$. Failure to ensure this would imply coordinate singularities in the analysis of the back reaction.
\\\\
We would next like to imagine that the shockwave emanated from the left boundary at a time $\tau_0>0$, moves along  
\be
U_0=e^{-\kappa \tau_0}.
\ee 
As the above Dray-'t Hooft solution is only valid for the shockwave travaelling along $U=0$ we have to ensure that we work in the limit $U_0\rightarrow 0$ which translates to the $\tau_0\rightarrow \infty$. In this limit we would like to determine the value of $\alpha$ $i.e.$ the magnitude of the back reaction induced on the metric. We do this following \cite{Shenker:2013pqa} by demanding the smoothness of the transverse volume density $H=\sqrt{{\rm det}[g_{\{\theta,z,\phi\}}]}$ along the shockwave
\be
H \big\vert_{U_0^+} = H\big\vert_{U_0^-}.
\label{smoothnes_cond_0}
\ee
Expanding this about the outer horizon we have
\be
H=H_0+H_1(r-\rpl)=H_0-H_1\left(\mathbb{A} \frac{U_0 V}{F'(\rpl)}\right)
\ee
where we express radial distances using the eq\eqref{r_to_UV}. The smoothness of transverse volume density across the shockwave thus implies
\be
H_0-H_1\left(\mathbb{A} \frac{U_0 V}{F'(\rpl)}\right)=\widetilde{H}_0-\widetilde{H}_1\left(\widetilde{\mathbb{A}} \frac{U_0 \widetilde{V}}{\widetilde{F}'(\rpl)}\right)\ee
where we use tilde to denote quantities after the shockwave $U>U_0$. To the leading order we can therefore find the shift in the $V$ coordinate to be
\bea
&&\widetilde{V}=\frac{\widetilde{F}'_+\mathbb{A}H_1}{F'_+\widetilde{\mathbb{A}}\widetilde{H}_1}V+\frac{(\widetilde{H}_0-H_0)\widetilde{F}'_+}{\mathbb{A}\widetilde{H}_1U_0},\hspace{0.3cm}{\rm where}\,\,F'_+\equiv F'(\rpl,\pi/2)
\label{Shift_V}
\eea
Note as $H_0$ is simply the transverse volume density at the outer horizon, integrating this along the periodic coordinates gives us the black hole entropy times $4G_N$. Therefore  we can write
\be
\widetilde{H}_0-H_0=
\delta H_0 \sim \beta(\delta \mathcal{M}-\mu\, (\delta J_{\phi_1}+\delta J_{\phi_2}))=\delta \mathcal{S}
\ee
where we use the ADM mass defined in AdS spaces and $\mu_{\phi_1}=\mu_{\phi_2}=\mu$
\be
\mathcal{M}=\frac{M}{\Xi^2},\hspace{0.3cm}J=\frac{Ma}{\Xi^2}
\label{ADM_M_J}
\ee
We would like to view the in-falling perturbation into the black hole as increasing the mass and angular momentum of the geometry and thus its entropy according to the first law. Moreover we would be interested in black holes with large entropy, in the limit of large entropy we observe that 
\be
\frac{F'_+}{H_1}=\frac{\mathbb{B}}{\mathcal{S}}
\ee
where $\mathbb{B}$ can be obtained by expanding $H$ and $F$ and is found to depend only on $\rmi/\rpl$ and $\theta$.
We imagine the shockwave to have very small energy as compared to the black hole mass, therefore the shockwave adds infinitesimally small entropy $\delta \mathcal{S}$ to the black hole where
\be
\frac{\delta \mathcal{S}}{\mathcal{S}}\rightarrow 0\hspace{0.2cm}\&\hspace{0.2cm}\delta J_{\phi_1}+\delta J_{\phi_2}=\Ll\,\delta M
\ee
Note we still can only work in the limit $U_0\rightarrow 0$, therefore we work in the scaling limit
\be
U_0\rightarrow 0,\hspace{0.3cm}\frac{(\delta\mathcal{M}-\mu\,(\delta J_{\phi_1}+\delta J_{\phi_2}))}{T_H \mathcal{S}}\rightarrow 0,\hspace{0.3cm}\frac{\delta\mathcal{M}(1-\mu\,(\Ll_{\phi_1}+\Ll_{\phi_2}))}{U_0\,T_H \,\mathcal{S}}\rightarrow{finite}.
\label{limit}
\ee
where we made use of the first law and the previous relation.
In such a limit the shift in $V$ would still be finite.
The parameters $(\mathbb{B},\mathbb{A},2\pi)$ are non-extensive and we can adsorb them  in the variation of ADM mass and angular momentum we can write
\be
\widetilde{V}=V-\frac{\Theta(U-U_0)}{U_0}\frac{\delta\mathcal{M}(1-\mu\,(\Ll_{\phi_1}+\Ll_{\phi_2}))}{T_H \mathcal{S}}
\label{V_shift_3}
\ee
where we have introduced the step function to indicate the shift in $V$ across the shock-wave.
\\\\
It is crucial to note  that the shift $\delta\mathcal{M}$ has a $\theta$ dependence as it  is  supposed to solve a differential equation implied by \eqref{backreaction_diff_axi_symmetric}.  This can be easily seen at $\delta H_0$ is simply the change in horizon area density which needs to be integrated along the periodic coordinates to give $\delta\mathcal{S}$. This reflects the fact that the geometry after the shockwave is no longer a stationary solution to Einstein's equations hence the $\delta\mathcal{M}$ can and must have spatial dependence\footnote{The time dependence comes $via$ the dependence on $V$ and would play an important role if one relaxes the limit $U_0\rightarrow 0$. }. Comparing this with  \eqref{V_shift_0} we see that 
\be
\alpha f(\theta)=\frac{\beta\,\delta\mathcal{M}(1-\mu\, (\Ll_{\phi_1}+\Ll_{\phi_2}))}{U_0\,\mathcal{S}}=\frac{\beta (1-\mu\,(\Ll_{\phi_1}+\Ll_{\phi_2}))E_0}{U_0\,\mathcal{S}}f(\theta)
\ee
where the transverse $\theta$ dependence on the right is captured by $\delta\mathcal{M}$. We have taken $\delta\mathcal{M}$ to be proportional to the total energy $E_0$ of the perturbation measured\footnote{The absorption of non-extensive constants such as $(\mathbb{B},\mathbb{A},2\pi)$ would not effect this identification in the final result.} at the boundary at  left exterior time $\tau_0$. Therefore we have
\be
V\rightarrow \widetilde{V}=V-\alpha f(\theta)\hspace{0.3cm}{\rm where }\,\,\,  \alpha=\frac{1}{U_0}\frac{\beta (1-\mu\,(\Ll_{\phi_1}+\Ll_{\phi_2}))E_0}{\mathcal{S}},\hspace{0.3cm} 
\label{V_shift}
\ee 
We can see that the shockwave grows exponentially as $U_0^{-1}=e^{\kappa \tau_0}$ $i.e.$ as $\tau_0$ tends towards infinity. The rate of growth of this exponential is governed by $\kappa$ for $(\Ll_{\phi_1}+\Ll_{\phi_2})< 1/\mu$. In what usually follows in finding a Dray-'t Hooft solution one finds the form of the response function $f(\theta)$ constrained by \eqref{backreaction_diff_axi_symmetric} which can be solved with recourse to spherical harmonics. For the case of single null particle falling in with a specific angular momentum $\Ll$ the response function would would be a function of $\{\theta,\phi,z\}$ and would  likewise be constrained by a sourced differential equation of the type \eqref{backreaction_diff_axi_symmetric} with the $rhs$ localizing the particle on $S^3$. In either cases the response function $f$ would determine how fast the backreaction spreads along the sphere and thus determine the `butterfly velocity' $v_B$. For the case at hand of an equatorial shockwave at every point in $\phi$ and $z$ it is apparent that the backreaction would be peaked at the equator and falloff gradually towards the poles in a manner dictated by \eqref{backreaction_diff_axi_symmetric}. Though interesting physics of the butterfly velocity $v_B$ can be discerned from such an analysis we would not concerning ourselves with it in this paper. We would therefore choose a normalization where $f(\theta=\pi/2)=1$ as we would only be concerned with dynamics of the metric at the equator.  
\\\\
Throughout this paper we assume that the shockwave emanates at the left $AdS_5$ boundary at a time $\tau_0$ at $\theta=\pi/2$ from every point in the $\{\phi,\psi\}$ plane. As is apparent we choose to work with the non radial coordinates $\{\tau,\theta,\phi,z\}$ given by \eqref{tau_to_t_phi_psi}\eqref{psi_to_z_tau} and \eqref{psi_to_z_coeff}. These coordinates are forced upon us by demanding that the components of the metric as seen along the infalling null rotating perturbation be smooth in the near horizon region, which is precisely the region in the bulk where the backreaction  is the strongest. Therefore in the Dary-'t Hooft analysis we have to demand that the response function is periodic in the $z$ coordinate\footnote{There is also a periodicity in the $\phi$ coordinate which remains the same.} and not the Boyer-Lindquist coordinate $\psi$. Assuming otherwise would lead to singularities in the metric components which end up in determining the differential equation \eqref{backreaction_diff_axi_symmetric} constraining the response function $f$. This can be understood as working with the co-moving coordinates once we are in the ergo-region as there are no possible stationary time like observers possible in this region. Therefore for the case of the backreaction due to a single in-falling null particle the equivalent shift in one of the Kruskal coordinates is given by $\alpha f(\theta,\phi,z)$ where $\alpha$ takes the form given above. Crucially, the exponential growth is only due to the presence of $U_0$.

This also implies that there is spread in time stamp associated with the start of the perturbation at the left asymptotic boundary as seen by a stationary boundary observer $i.e.$ the proper time $t^{(s)}$ of the dual CFT given by \eqref{tau_to_t_phi_psi}   
\be
\tau_0=t_0^{(s)} + \frac{\Ll_{\phi_1}-\Ll_{\phi_2}}{2}\phi_0-\frac{\Ll_{\phi_1}+\Ll_{\phi_2}}{2}\psi_0.
\label{tau0_to_ts0_phi0_psi0}
\ee
Given that $z$ is defined such that $\psi\in[0,2\pi]$ implies $z\in[0,2\pi]$ in the near horizon region\footnote{This is fixed by by demanding $\eta$ take the specific value in \eqref{psi_to_z_coeff}.} the above relation gives the spread in $t_0^{(s)}$ for a fixed $\tau_0$ for shockwave originating at evry point in the $\{\phi,\psi\}$ plane. In what follows we will only deal with $\tau_0$ as working with $t_0^{(s)}$ only produces a shift which would not effect the results qualitatively. 
\\\\
Before we proceed to compute the change in the extremal surface due the above shockwave geometry it would be useful to understand how the shockwave travels within the (left) exterior of the geometry given the limits \eqref{limit}. The Kruskal coordinates used helped us to write down the solution to the shockwave present along the past horizon, however this is used as a limiting case to the scenario where the shockwave emanates at $U_0\rightarrow 0$ from the left $AdS$ boundary. The limit \eqref{limit} can be seen to define how close the shockwave is to the past horizon ($i.e.$ as close as $\delta\cS/\cS$ is to $0$ in terms of $U_0$.). The shockwave intersects the future horizon at $U=U_0$ and goes past it depending on the turning point behaviour discussed in subsection-3.3.1. Upon reaching the turning point the shockwave reverses direction (in terms of its radial coordinate) and begins to travel along $V=const$. The location of the turning point requires further numerical analysis ($c.f.$ subsection-3.3.1) and can be behind even the inner horizon. However this would not be of concern to us as we would dealing with the change in $I(A:B)$ up until the time the shockwave just reaches the outer horizon at $t=0$. This above scenario is similar to the one studied recently in \cite{Horowitz:2022ptw} where a charged non-rotating shockwave was analysed in RN $AdS_4$ extending the analysis of \cite{Leichenauer:2014nxa}. We will therefore see a similar effect as seen in \cite{Horowitz:2022ptw} exhibiting a delayed onset of chaos.  
\section{Extremal surface along the equator}
In this section, we determine the effect of shockwave on the holographic mutual information between subsystems in the left and right CFT. The holographic mutual information involves three terms each of which corresponds to the holographic entanglement entropy of a subsystem and may be computed through the  HRT proposal as follows
\begin{align}\label{IAB}
	I(A:B)&=S_A+S_B-S_{AB}\nonumber\\
	S_x&=\frac{{\cal A}_{\gamma_x}}{4G_N^{(5)}}
\end{align}
where ${\cal A}_{\gamma_x}$ corresponds to the area of a co-dimension two extremal surface $\gamma_x$ homologous to $x$ and $G_N^{(5)}$ is the five dimensional Newtons's gravitational constant. The areas of the extremal surfaces and the corresponding holographic entanglement entropies for subsystems $A$ and $B$ are unaffected by the shockwave at late times as they are sufficiently away from the black hole horizon. Here, as mentioned in the introduction the subsystems $A$ -in the $left$ $CFT$, \& $B$ in the $right$ $CFT$, are hemispheres ending on the equator described by $\theta=\pi/2$. Therefore the HRT surface homologous to $A\cup B$ always lies on the equatorial plane $\theta=\pi/2$, this can be easily seen by the vanishing of the extrinsic curvature of any co-dim 2 equatorial surface along the $\theta$ direction. We further need to extremize the co-dim 2 surface by moving it in the time direction. Given that the remaining non-radial spatial coordinates are symmetric directions of the black hole and also of the back-reacted geometry\footnote{This is because a shockwave is not localized at any values of $\phi$ and $z$.}, this effectively involves extrimizing  a space-like length in a particular 2d metric along the $r$ and $\tau$ directions.
\\\\
In order to perform this computation we utilize the method described in \cite{Leichenauer:2014nxa} to estimate the growth of the area of the extremal surface which the present authors utilized in \cite{Malvimat:2022oue} to obtain the change in holographic mutual information of the subsystems in dual CFTs due to shockwaves in Kerr-$AdS_4$ spacetime. The induced metric in equatorial plane is given by
\be
ds_{\theta=\frac{\pi}{2}}^2=F\left(-d \tau^{2}+\frac{d r^{2}}{f^{2}}\right) + h_z(dz+g_\phi d\phi+A_\tau d\tau)^2 +h_\phi(d\phi+h_\tau d\tau)^2  
\label{Inducedmetric_kruskal_L}
\ee
The area of the required extremal surface corresponding to the holographic entanglement entropy $S_{AB}$ may be expressed as follows
\begin{align}\label{area1}
{\cal A}_{\theta=\frac{\pi}{2}}
&=S_{z}S_{\phi} \int d \tau \sqrt{h_z h_\phi} \sqrt{-F+F f^{-2} \dot{r}^{2}}
\end{align}
Notice that the integrand is independent of the $\tau$ coordinate which leads to a conserved quantity $\gamma$ expressed as follows
\begin{align}
\gamma=\frac{-F \sqrt{h_z h_\phi}}{\sqrt{-F+F f^{-2} \dot{r}^{2}}}=\sqrt{-F_{0}h_{z_0} h_{\phi_0}}. 
\end{align}
where $F_{0},h_{z_0}, h_{\phi_0}$ are the respective values of the functions $F,h_z,h_\phi$ where $\dot{r}=0$.  We may invert the above equation to obtain an expression for $\tau(r)$ as follows
\begin{align}\label{tauint}
	\tau(r)=\int d r \frac{1}{f \sqrt{1+\gamma^{-2} F h_z h_\phi}}
\end{align}

The area may now be computed by dividing it into two symmetric halves and separately evaluating three different segments of each half. The first part is described by a segment which starts from boundary and ends at $(U_a,0)$ the intersection of the extremal surface with $V=0$. The second segment begins at ($U_a,0)$ and ends at  $(U_b,V_b)$ which is a constant $r=r_0$ surface where $r_0$ corresponds to the turning point of the extremal surface whereas the third segment begins from  $(U_b,V_b)$ and terminates at the radial point  $(0,\alpha/2)$ which is the other horizon. We will not describe the full computational details here as it  is comprehensively explained in \cite{Leichenauer:2014nxa,Malvimat:2022oue}. Basically we obtain the following expression for $alpha$ which involves three different integrals $G_1,G_2,G_3$ as follows
\begin{align}\label{alphaint}
	\alpha&=2 \exp \left(G_{1}+G_{2}+G_{3}\right)\\ \textrm{where,} &\nonumber\\
\,  \, \quad \quad  G_{1} &=-2 \kappa \int_{\bar{r}}^{r_{0}} \frac{d r}{-f} \\ G_{2} &=2 \kappa \int_{r_{0}}^{\infty} \frac{d r}{f}\left(1-\frac{1}{\sqrt{1+\gamma^{-2} F h_z h_\phi  }}\right) \\ G_{3} &=2 \kappa \int_{r_{0}}^{r_{+}} \frac{d r}{f}\left(1+\frac{1}{\sqrt{1+\gamma^{-2} F  h_z h_\phi }}\right)\label{G3} 
\end{align}
where $\bar{r}$ corresponds to the point where $r_*$ vanishes inside the horizon i.e $r_*(\bar{r})=0$.
Note that the first two integrals given by $G_1,G_2$ diverge only as $r_0\to r_+$ which also corresponds to the limit in which the shift parameter $\alpha \to 0$ whereas the third integral diverges only as $r_0\to r_c$, $r_c$ being the solution to the following equation
\begin{align}
	F'(r_c)  h_z(r_c) h_\phi(r_c)+	F(r_c)  h'_z(r_c) h_\phi(r_c)+	F(r_c)  h_z(r_c) h'_\phi(r_c)=0
\end{align}
Therefore as $r_0\to r_c$ the $G_3$ integral diverges which in turn implies that $\alpha \to \infty$ from eq.\eqref{alphaint}. By utilizing this result we may compute the leading contribution to the area
integral in eq.\eqref{area1}. To this end we may first utilize eq.\eqref{tauint} in eq.\eqref{area1} to obtain the following expression for the area integral
\begin{align}\label{area2}
 \mathcal{A}_{\theta=\frac{\pi}{2}}=8\pi^2\int \frac{d r}{f} \frac{F h_z h_\phi / \gamma}{\sqrt{1+\gamma^{-2} F h_zh_\phi}}
\end{align}
Here we used $S_z=4\pi$ and $S_\phi=2\pi$.
We may now re-express the above area integral  by expanding around $r=r_c$  in terms of the divergent integral $G_3$ in eq.\eqref{G3}.
Note that the required extremal surface gets dominant contribution from four times the area of the segment  as expressed below 
\begin{align}
		\mathcal{A}_{\theta=\frac{\pi}{2}}&\approx8\pi^2\sqrt{-F_c \, h^c_z \, h^c_\phi}\frac{G_3}{2\kappa} \label{area3}\\
	\mathcal{A}_{A\cup B}	=4\mathcal{A}_{\theta=\frac{\pi}{2}}&\approx\frac{16\pi^2}{\kappa}\sqrt{-F_c \, h^c_z \, h^c_\phi}\log[\alpha] \label{area4}
\end{align}
where we have also utilized the relation between $\alpha$ and $G_3$ in eq.\eqref{alphaint}. Having obtained the area of the required HRT surface we may now substitute the expression for $\alpha$ given by eq.\eqref{V_shift} in the above equation to obtain
\begin{align}\label{areaAB}
		\mathcal{A}_{A\cup B}	\approx\ 16\pi^2\tau_0\sqrt{-F_c \, h^c_z \, h^c_\phi}+ \frac{16\pi^2}{\kappa}\log[\frac{\beta E_0(1-\mu\,(\Ll_{\phi_1}+\Ll_{\phi_2}))}{\mathcal{S}}]
\end{align}
\begin{figure}
	\centering
	\includegraphics[scale=0.2]
	{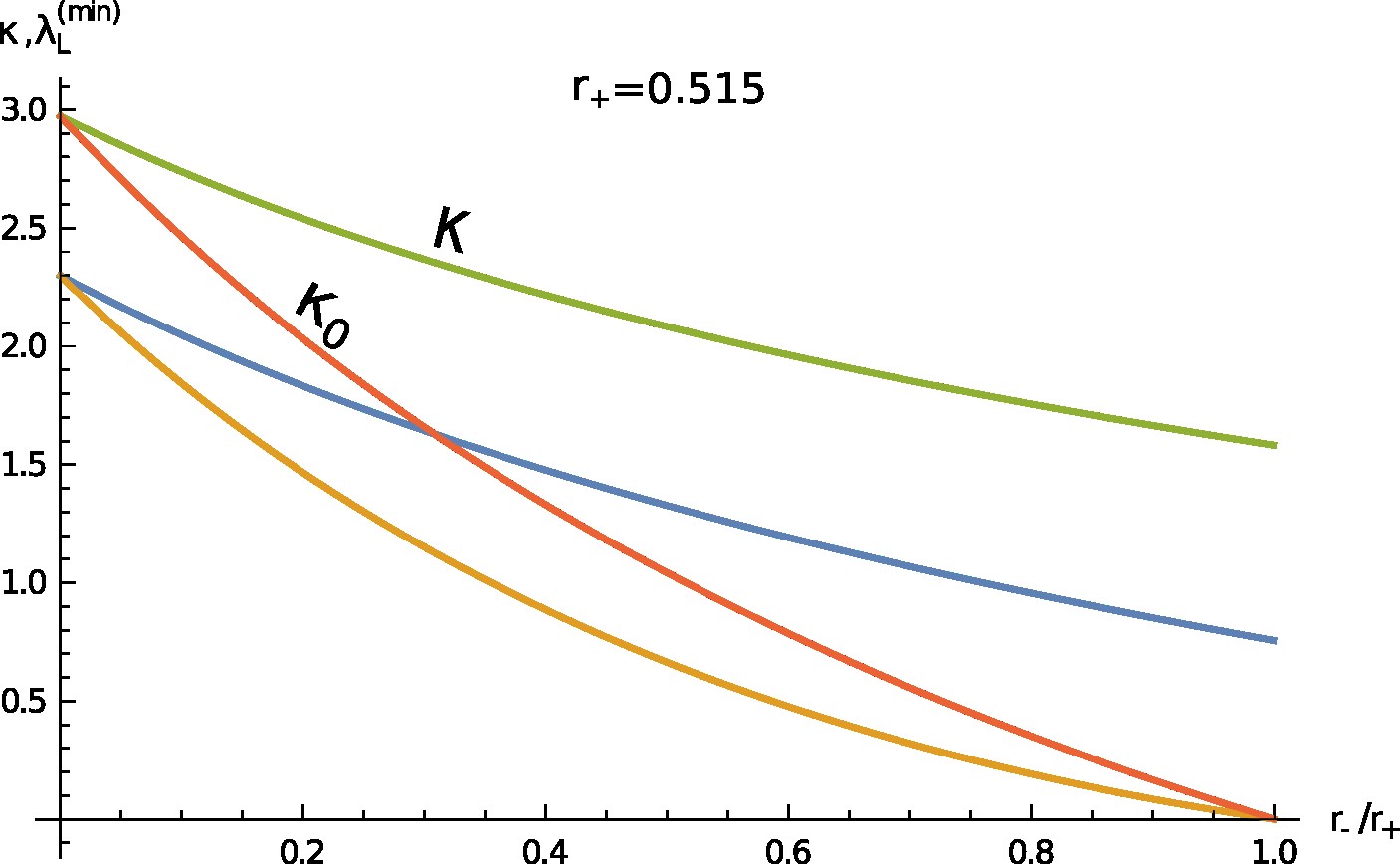}
	\caption{Plots for $\lambda_L^{(min)}$(blue,orange) with and without angular momentum along with that for $\kappa$ (green) \& $\kappa_0=2\pi T_H$ (red) against $\rmi/\rpl$. Here $\Ll_+=\frac{\rmi}{\rpl}\mu^{-1}$ and $\Ll_-=\varepsilon\Ll_+$ for the green curve and the blue curve corresponding to $\lambda_L^{(min)}$ with angular momentum. We have taken $\ell=1$. The plot is found to independent of $\varepsilon$. The bottommost (orange) curve is the plot for  $\lambda_L^{(min)}$ when $\Ll_\pm=0$ $i.e.$ when the shockwave has no angular momentum. The above values of $\Ll_\pm$ are allowed by turning point analysis except for for very small values of $\rmi/\rpl$ for the blue and orange  plots. It is quite apparent that the instantaneous $\lambda_L^{(min)}$ can be greater than $\kappa_0$ for non-extremal geometries. Here we exclude the extremal point $\frac{\rmi}{\rpl}=1$.  }
\label{F_cH_c_kappa}
\end{figure}
This completes the computation of the growth of extremal surface due to the perturbation. We now use the above expression to approximate the holographic mutual information $I(A:B)$ through the application of the $RT/HRT$ formula as described by eq.\eqref{IAB}. 
We would like to have some idea about the instantaneous Lyapunov exponent $\lambda_L$ as the shockwave affects the geometry at very late times. As we have seen at such time scales $I(A:B)$ changes only due to change in entanglement entropy $S_{A\cup B}$ as the HRT surface corresponding to it traverses the horizons in the Kruskal diagram. Therefore the instantaneous $\lambda_L$ at late times can be discerned by knowing the rate of growth of the perturbation in $S_{A\cup B}$ due to the shockwave as compared to it's unperturbed value. However in geometries such as Kerr the HRT/RT surfaces corresponding to $S_{A\cup B}$ are difficult to compute. In such cases we can estimate the magnitude of $S_{A\cup B}$ by the entropy of the black hole $\mathcal{S}$. The understanding being that for purification of large enough systems the entanglement entropy roughly peaks when one integrates out half of the degrees of freedom\footnote{The entropy of the black hole $\mathcal{S}$ can be thought of as obtained by integrating out one of the 2 CFTs corresponding to the TFD description.}. Therefore for the unperturbed value of $S^{(0)}_{A\cup B}=\mathcal{A}^{(0)}_{A\cup B}/(4G_N)$ we can write  
\be
\mathcal{A}_{A\cup B}^{(0)}=\rho \mathcal{A}_H,\hspace{0.3cm}\rho\in \{0,1\}
\label{area_AB_unchanged}.
\ee
Defining the instantaneous Lyapunov index as
\be
\mathcal{A}_{A\cup B}=\lambda_L\tau_0\mathcal{A}^{(0)}_{A\cup B}
\label{lambda_def}
\ee
we find the minimum value of $\lambda_L$ for $\rho=1$ as
\be
\lambda^{(min)}_L=\frac{\mathcal{A}_{A\cup B}}{\mathcal{A}_H}=\frac{16\pi^2\sqrt{-F_ch^c_zh_\phi^c}}{\mathcal{A}_H}.
\label{lambda_min}
\ee 
\begin{figure}
	\centering
	\includegraphics[scale=0.2]
	{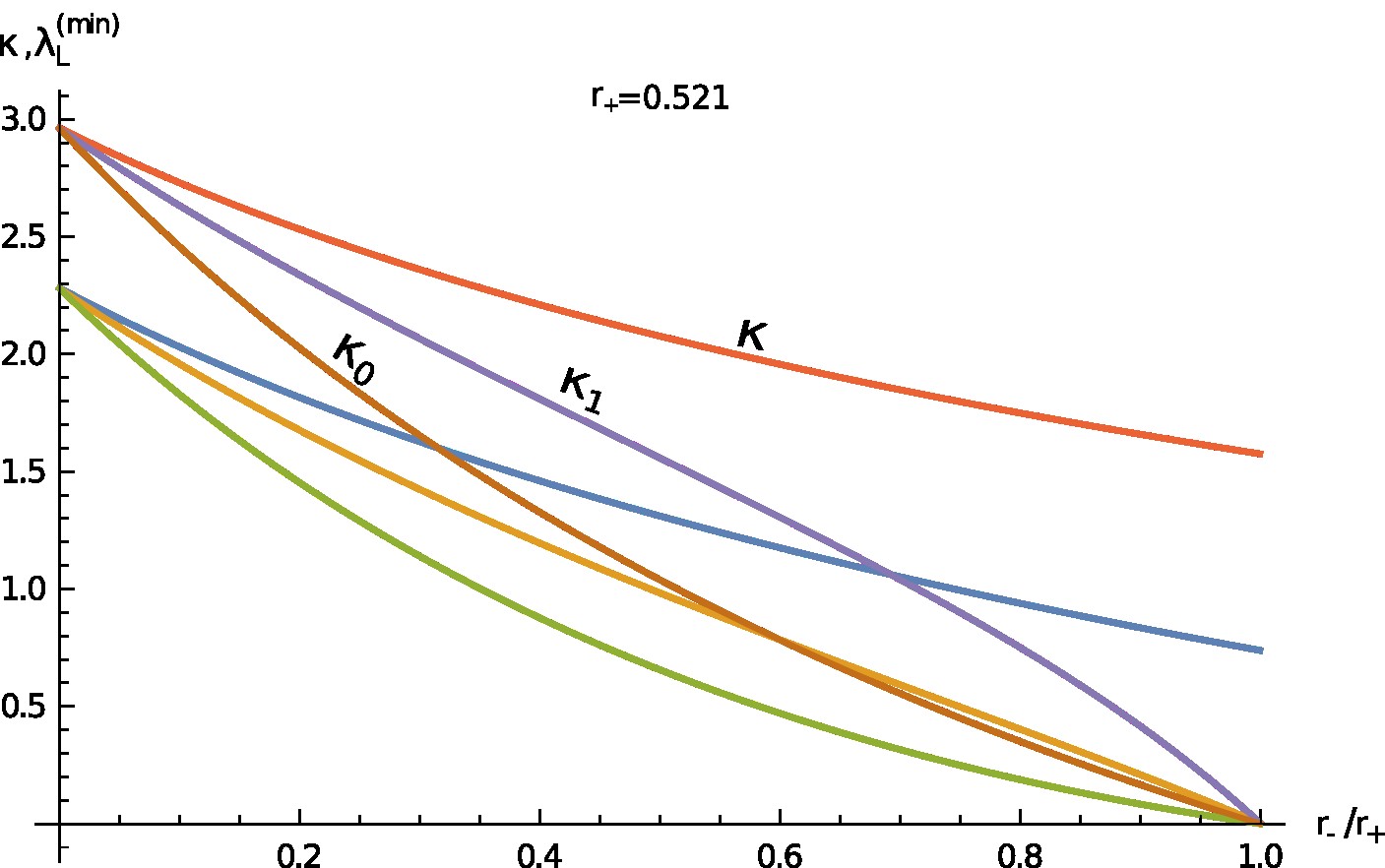}
	\caption{Plots various $\lambda_L^{(min)}$(blue,orange,green) and blueshift exponents $\kappa$(red,violet,brown) for different values of $\Ll_+<\mu^{-1}$. Here $\rpl=0.521$ and $\ell=1$. The red plot labelled $\kappa$ and the blue plot for $\lambda_L^{(min)}$ have $\Ll_+=\frac{\rmi}{\rpl}\mu^{-1}$. The violet plot labelled $\kappa_1$ and the orange plot for $\lambda_L^{(min)}$ have  $\Ll_+=\frac{2}{3}\frac{\rmi}{\rpl}\mu^{-1}$. The brown plot labelled $\kappa_0$ and the green plot for $\lambda_L^{(min)}$ have $\Ll_+=0$ $i.e.$ $\kappa_0=2\pi T_H$. $\Ll_-=0$ for all the plots above as it seen not to affect the functional behaviour. It is apparent that the blue plot for $\lambda_L^{(min)}$ is only bounded by the plot labelled kappa(red) and not by $\kappa_1$ or $\kappa_0$. Similarly the orange plot for $\lambda_L^{(min)}$ is not bounded by $\kappa_0$ but by those labelled $\kappa_1$ and $\kappa$. Here the plots exclude the extremal point $\frac{\rmi}{\rpl}=1$.}
\label{F_cH_c_kappa_1}
\end{figure}
This is the minimum value of the instantaneous Lyapunov index attained at very late times, the actual value can be greater than this. We plot\footnote{Unlike the case in Kerr $AdS_4$ here the value of $r_c$ and consequently the minimum value of $F_ch^c_zh^c_\phi$ is determined numerically.} this $\lambda_L^{(min)}$ against $\rmi/\rpl$ for $\ell=1$ in Fig-\ref{F_cH_c_kappa} along with the values for $\kappa$ and $\kappa_0=2\pi T_H$. Here the sum of the angular momenta of the shockwave scales as $\Ll_+=\Llo+\Llt=\frac{\rmi}{\rpl}\mu^{-1}$. The orange plot is for the case when the shockwave has no angular momenta along any of the directions, we see that this is clearly bounded by $\kappa_0$ and goes to zero as $\rmi/\rpl\rightarrow 1$. However the blue curve for $\lambda_L^{(min)}$ for $\Ll_+=\frac{\rmi}{\rpl}\mu^{-1}$ is clearly greater than $\kappa_0$ but is always bounded by $\kappa$. It is interesting to note that $\lambda_L^{(min)}$ does not seem to depend on $\Ll_-$ while the function $Fh_zh_\phi$ in general depends on $\Ll_\pm$. 
\begin{figure}
	\centering
	\includegraphics[scale=0.18]
	{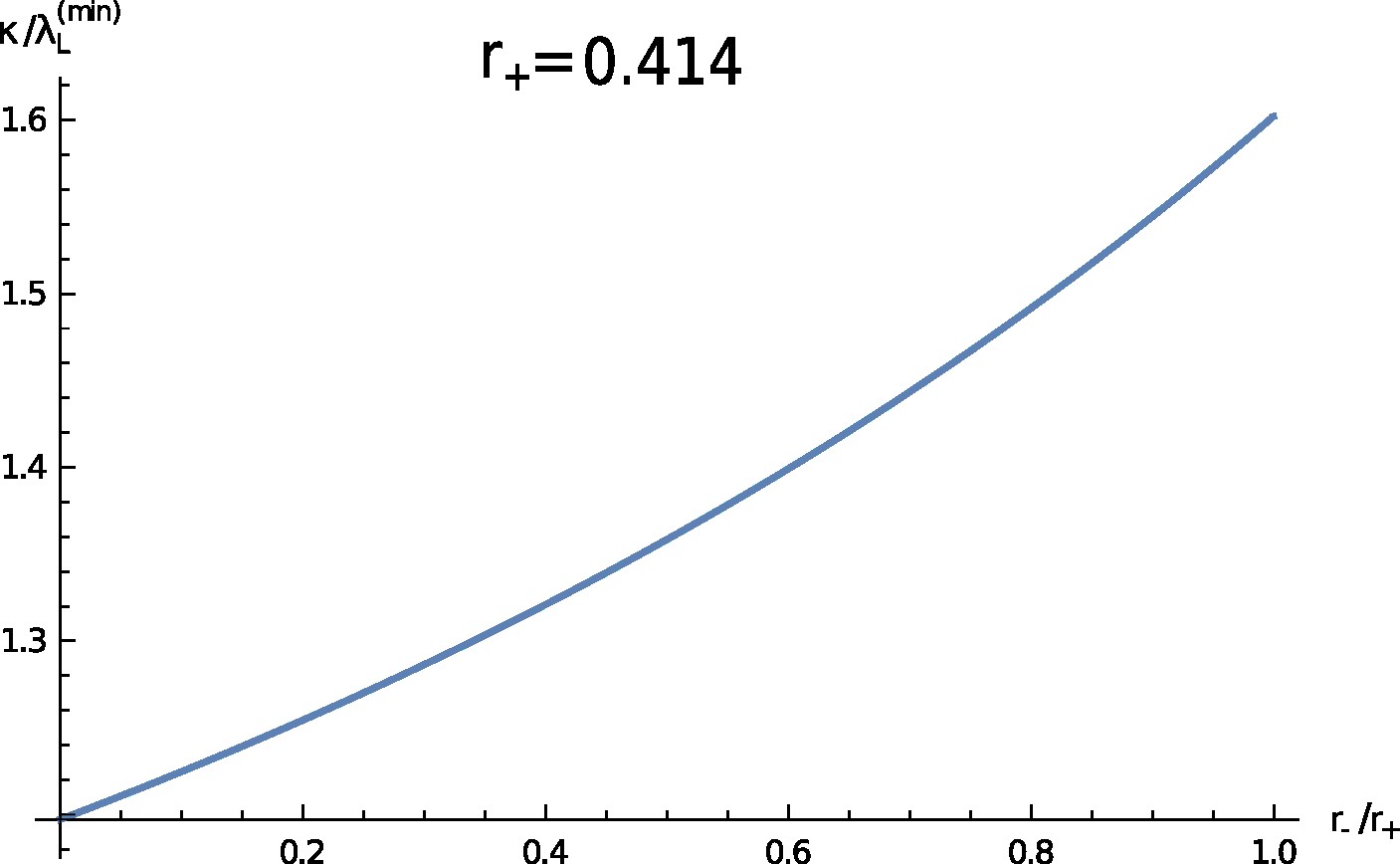}
	\caption{Plot of $\kappa/\lambda_L^{(min)}$ with $\ell=1$ against $\rmi/\rpl$. This plot is independent of both $\Ll_+$ and $\Ll_-$ and depends only on the black hole outer horizon radius $\rpl$.}
\label{kappa_by_lambda_L_min}
\end{figure}
\\\\
To get a better idea on the dependence of $\lambda_L^{(min)}$ on $\Ll_+$ we plot in Fig-\ref{F_cH_c_kappa_1} the blueshift exponents $\kappa$ and $\lambda_L^{(min)}$ with $\Ll_+=\varepsilon_+\frac{\rmi}{\rpl	}\mu^{-1}$ for $\varepsilon_+\in\{1,\frac{2}{3},0\}$. We find the  $\lambda_L^{(min)}$ for a particular value of $\varepsilon_+$ is effectively bounded by the $\kappa$ for the same value of $\varepsilon_+$ or higher. One can plot $\kappa$ and $\lambda_L^{(min)}$ against $\rmi/\rpl$ for larger number of fractions between $\varepsilon_+\in [1,0]$ and reach the same conclusion. We also find empirically that $\kappa/\lambda_L^{(min)}$ does not depend on $\Ll_+$ and approaches $1$ for small values of the outer horizon.
It  is physically understandable that $\lambda_L$ be bounded by the blueshift suffered by the shockwave in each of the cases above as the later determines the rate at which the metric changes. However since these are the plots for $\lambda_L^{(min)}$ the actual value of instantaneous $\lambda_L$ can be greater than the respective $\kappa$. 
\\\\
There is another way one can discern $\lambda_L$, by determining the scrambling time $\tau_*$. In this case it is the time required by the shockwave to completely disrupt the mutual information between the subsystems $A$ and $B$. 
Substituting the area of the extremal surface homologous to the subsystem-$AB$ given by eq.\eqref{areaAB} in \eqref{IAB} for the holographic mutual information we obtain
\begin{align}
I(A:B)\approx S_A+S_B-\ \frac{16\pi^2\tau_0}{G_N}\sqrt{-F_c \, h^c_z \, h^c_\phi}- \frac{16\pi^2}{\kappa G_N}\log[\frac{\beta E_0(1-\mu\,(\Ll_{\phi_1}+\Ll_{\phi_2}))}{\mathcal{S}}]
\end{align}
We may now utilize the above expression to find out the scrambling time $\tau_0=\tau_*$ at which the holographic mutual information vanishes. This leads us to
\begin{align}\label{scrambling_time}
&\kappa \tau_*\approx\log\mathcal{S}+\log\bigg[\frac{1}{(1-\mu\,\Ll_+)}\bigg]+\frac{\kappa(\mathcal{A}_A+\mathcal{A}_B)}{16\pi^2\sqrt{-F_c \, h^c_z \, h^c_\phi}},\\
&{\rm where} \hspace{0.5cm} \kappa=\frac{2\pi T_H}{(1-\mu\,\Ll_+)},\,\,\,\,\Ll_+=\Llo+\Llt
\label{kappa_fin}
\end{align}
where we have take $\beta E_0\approx 1$ implying that the shockwave has energy as measured at the $AdS$ boundary to be of the order of the temperature of the black hole. The last term in \eqref{scrambling_time} can be written using \eqref{lambda_min} as 
\be
\frac{\kappa(\mathcal{A}_A+\mathcal{A}_B)}{16\pi^2\sqrt{-F_c \, h^c_z \, h^c_\phi}}=\frac{\kappa(\mathcal{A}_A+\mathcal{A}_B)}{\lambda^{(min)}_L\mathcal{A}_H}
\label{3rd_term}
\ee
which we for large enough subsystems $A$ and $B$ can be seen not to scale with the entropy of the system. Plotting $\kappa/\lambda_L^{(min)}$ against $\rmi/\rpl$ for various values of $\rpl$ we find that this is independent of $\Ll_+=\Llo+\Llt$ see Fig-\ref{kappa_by_lambda_L_min}. Therefore this term exists as it is for even when the shockwave has no angular momentum in the axi-symmetry directions corresponding to the exponential blueshift with index $\kappa_0=2\pi T_H$. Further \eqref{3rd_term} can be regarded as
\be
\kappa \frac{S_A+S_B}{S_{A\cup B}}
\ee
and for large enough thermal systems like that of a large black hole each of $S_A$, $S_B$ and $S_{A\cup B}$
scale as some fraction of the black hole entropy $\mathcal{S}$.  Therefore the last term in \eqref{scrambling_time} depends on black hole parameters like $\rmi/\rpl$ only and not extensively on $\mathcal{S}$ or on details of the shockwave such as its angular momenta.
We thus find that at very late times leading up to the scrambling time the Lyapunov index is basically
\be
\lambda_L=\kappa=\frac{2\pi T_H}{(1-\mu\,\Ll_+)}
\label{lambda_large}
\ee 
which can be regarded as its large time average value, where as the $\lambda_L^{(min)}$ can be regarded as the minimum value of $\lambda_L$ at large times leading up to the scrambling time given above. 
\\\\
The second term in \eqref{scrambling_time} is the only term which depends on $\Ll_+$ and the fact that the geometry is Kerr via $\mu$. Although this term does not scale with $\mathcal{S}$ for large black holes like the last term in \eqref{scrambling_time}, it does however depend on $\Ll_+$ and is positive as $\Ll_+<\mu^{-1}$. The origin of this term can be traced back to the first law of black hole dynamics used in the previous section $c.f.$ \eqref{limit}. This term accounts for the delay in the onset of scrambling due to $\Ll_+$ and is also found to exist in scrambling times recently computed for Reisnner-Nordst\"{o}m black holes \cite{Horowitz:2022ptw}. Here for the case of a charged shockwave falling into the RN black hole the authors find a delay in the onset of scrambling proportional to
\be
t^{(q)}_*-t_*=\frac{1}{2\pi T_H}\log\left(1-\mu\frac{q}{E_0}\right)^{-1}
\ee
where $q/E_0$ is the charge per unit energy of the shockwave and $\mu$ is the chemical potential associated with the RN charge and $t_*$ is the scrambling time measured for an uncharged shockwave. As is apparent from the analysis in section 3 that this term is accounted for by the thermodynamics of black holes it is immediately clear that if the charge of the shockwave is reversed $q\rightarrow-q$ this would lead to a corresponding advance in the scrambling time $i.e.$ $t_*^{(q)}<t_*$. 
\section{Conclusion \& Discussions}
In this paper we study the butterfly effect in Kerr $AdS_5$ with equal angular momentum by analysing the disruption caused to the finely tuned mutual information $I(A:B)$ between large subsystems of the $left$ and $right$ CFTs due to in-falling rotating shockwaves. 
Both the shockwave and the subsystems are such that they respect the axi-symmetry of the geometry along periodic angular coordinates $\phi_1$ and $\phi_2$. 
Further like in the analysis of the butterfly effect in Kerr $AdS_4$ \cite{Malvimat:2022oue} we choose the shockwave to exist for every value of $\phi_1$ and $\phi_2$ and only confined to the equator. 
For late times $>>T_H^{-1}$ leading up to the scrambling time we are able to find the minimum value for the instantaneous Lyapunov index $\lambda_L^{(min)}$. 
This is basically the instantaneous minimum rate of growth of the HRT surface homologous to $A\cup B$ and traversing the Kruskal diagram. 
We plot this for different values of the shockwave's angular momentum $\Llo$ \& $\Llt$ Fig-\ref{F_cH_c_kappa},\ref{F_cH_c_kappa_1}. We find $\lambda_L^{(min)}$ to be bounded by $\kappa$ \eqref{kappa_fin} and independent of the $\Ll_-=\Llo-\Llt$. It is easily seen that $\lambda_L^{(min)}>2\pi T_H$ for sufficiently non-extremal geometries and can survive the limit to extremality in certain cases.
We also find that the exponent of the blueshift suffered by such a shockwave by the time it reaches the outer horizon from the $AdS_5$ boundary is given by $\kappa$ with $\kappa=\kappa_0=2\pi T_H$ for $\Ll_+=\Llo+\Llt=0$. Further we find that the ratio $\kappa/\lambda_L^{(min)}$ to be independent of $\Ll_\pm$. We are also able to estimate the scrambling time $i.e.$ time taken for the mutual information to go to zero \eqref{scrambling_time} and the late time Lyapunov exponent is given by $\kappa$ i.e. $\kappa\tau_*\sim\log \mathcal{S}$ where $\mathcal{S}$ is the black hole's  entropy. We also find corrections to this scrambling time which is positive and proportional to $\log(1-\mu\,\Ll_+)^{-1}$ causing a delay in the onset of scrambling, however this term is not extensive as $\kappa^{-1}\log\mathcal{S}$ and does not scale with the black hole's entropy. A similar term has also recently appeared in the analysis of RN black holes in $AdS_4$ \cite{Horowitz:2022ptw} where the phenomena of delay in the scrambling times due charged shockwaves have been investigated.  
\\\\
The analysis presented here corroborates a similar one presented for Kerr black holes in $AdS_4$ \cite{Malvimat:2022oue}. 
It is also interesting to contrast it with the similar analysis done for rotating shockwaves in BTZ \cite{Malvimat:2021itk} in which $\lambda_L$ was found to be greater than $2\pi T_H$ only for certain specific values of chemical potential $\mu$, shockwave angular momenta $\Ll$ and subsystems size. Analytic result was only obtainable for $\Ll=1$ where for $\mu>1/2$ we found $\lambda_L=\pi T_H/(1-\mu\,\Ll)>2\pi T_H$. 
For the case of Kerr $AdS_4$ and $AdS_5$ we see  a much more cleaner result in terms of the behaviour of the instantaneous $\lambda_L^{(min)}$ and the scrambling time. This may be due to higher dimensions allowing a choice of subsystems and HRT surfaces in the bulk which respect the rotational(axial) symmetry of the geometry. 
\\\\
It has been well known that the low energy chaotic behaviour of large$-N$ holographic systems is captured by the near horizon dynamics and effective theory of black holes. 
The JT theory of gravity has been shown  to reproduce the near extremal thermodynamics of wide class of black holes \cite{Moitra:2018jqs,Castro:2018ffi} for fluctuations in mass and entropy over extremality with the other black hole charges held fixed.
Such a JT theory can be used to deduce the $\lambda_L$ due to interactions with 2d probe scalar fields in the near horizon region and is given by $\lambda_L=2\pi T_H$ \cite{Jensen:2016pah,Maldacena:2016upp}. However the phenomena explored in this paper (and in \cite{Malvimat:2022oue,Malvimat:2021itk}) concerns with perturbing a Kerr black hole with $\delta M$ and $\delta (J_{\phi_1}+J_{\phi_2})=\Ll_+\delta M$. The first law in this case takes the form
\be
\delta M=\frac{T_H}{(1-\mu \,\Ll_+)}\delta \mathcal{S}
\label{first_law}
\ee
as $\mu_1=\mu_2=\mu$. It would be interesting to see if a JT like effective theory of gravity can capture the thermodynamics of  such departures form extremality Kerr black holes. This question has recently been answered for the case of near extremal BTZ geometries where such departures form extremality are also seen to be governed by the JT theory \cite{Poojary:2022meo}. Here the near horizon geometry and the JT theory is parametrized by specific angular momentum $\Ll$ used to obtain the null ingoing geodesics, the near horizon limit is then taken along such null geodesics. Such a JT description captures the relevant near extremal thermodynamics and the temperature $T_H^{(2)}$ of the $AdS_2$ geometry in the JT theory is related to the near extremal BTZ temperature $T_H$ as
\be
T^{(2)}_H=\frac{T_H}{1-\mu\,\Ll}
\label{JT_tilde_BTZ}
\ee   
It would therefore be interesting to obtain similar effective theories in the near horizon region of near extremal Kerr black holes in dimensions greater than 3 \cite{Poojary:2022vsz}. It must also be noted that as far as thermodynamics is concerned the JT description for Kerr black holes as in \cite{Moitra:2018jqs} captures the behaviour in an ensemble where angular momenta $J_{\phi_1,\phi_2}$ are held fixed. 
The perturbation studied in this work would require one to work  in a canonical ensemble where $M-\Llo J_{\phi_1}$ and $M-\Llt J_{\phi_2}$ are held fixed instead. 
\\\\
The work of Castro $et\, al$ \cite{Castro:2021fhc} explores the near horizon effective action for near extremal Myers-Perry type black hole in $AdS_5$ and studies the corrections obtained to the (dynamical part of the) JT theory in the near horizon and near extremal limit. The above Lyapunov exponent \eqref{lambda_large} could therefore probably be obtained from the near horizon picture used in \cite{Castro:2018ffi,Moitra:2018jqs}  if such corrections are sufficiently accounted for in the effective graviton exchanges between probe matter fields. 
Given that sufficiently late time dynamics of perturbation is able to see a $\lambda_L$ given by \eqref{lambda_large}, its quite probable that the deep IR behaviour of probe matter in near extremal Kerr geometries in $AdS_{5}$ is not simply obtainable by the JT model as studied in \cite{Castro:2018ffi,Moitra:2018jqs,Castro:2021fhc}.
\section*{Acknowledgements}
RP would like to thank Daniel Grumiller and  Prashant Kocherlakota for discussions pertaining to some aspects of this work. RP is supported by the Lise Meitner grant M-2882 N of the FWF. 
\bibliographystyle{JHEP.bst}
\bibliography{bulk_syk_soft_modes.bib}
\end{document}